\documentclass[aps,prl,reprint,groupedaddress]{revtex4-2}
\usepackage{commath}
\usepackage{braket}
\usepackage{hyperref}
\usepackage{dsfont}

\begin{document}

\title{T-duality and bosonization as examples of continuum gauging and disentangling}

\author{Gertian Roose}
\affiliation{School of Physics and Astronomy, Tel Aviv University, Tel Aviv 6997801, Israel}
\email{Gertian.Roose@gmail.com}

\author{Erez Zohar}
\affiliation{School of Physics and Astronomy, Tel Aviv University, Tel Aviv 6997801, Israel}

\date{\today}

\begin{abstract}
Dualities and duality transformations form a well established methodology in various aspects of quantum many body physics and quantum field theories, allowing one to exploit equivalence between models which may naively seem completely different in order to gain access to further physical regimes, either analytically, numerically or experimentally. Recently, in the context of condensed matter physics and quantum information, it was shown that dualities can be understood very well through 
 a gauging and disentangling procedure that can be represented by a finite depth quantum circuit. 
 In this letter we expand these concepts to the continuum, suggesting them as a way to derive duality transformations in continuum field theories and particle physics, and benchmark the presented ideas through the re-derivation of T-duality and bosonization.
\end{abstract}

\maketitle

For a long time, dualities \cite{PhysRev.60.252,Fradkin1978,Savit1980,Cobanera2010,Cobanera2011} have been an intriguing and meaningful property of physical theories. There can be different meanings and definitions of duality, but in general, if two physical models are dual, it implies that they are both equivalent mathematical descriptions of the same observable physics, either classical or quantum. Lattice examples include dualities of Ising models, lattice gauge theories and similar models (e.g. \cite{PhysRev.60.252,Fradkin1978,Savit1980,Elitzur1979,Horn1979,Ortiz2012}), or mapping between fermionic degrees of freedom to spin-like objects (e.g. \cite{Ball2005,Verstraete2005,Kitaev2006,Nussinov2009,Nussinov2012,Zohar_2018}).

More modern examples are, for example,  AdS/CFT correspondence \cite{Maldacena:1997re}, Seiberg dualities \cite{Seiberg:1994pq}, T-duality \cite{Alvarez:1996up, BUSCHER1988466}, bosonization \cite{Fradkin_Field_Theories_of_Condensed_Matter_Physics,von_Delft_1998} and particle-vortex duality \cite{PESKIN1978122, PhysRevLett.47.1556}. Since strongly coupled theories are often dual to weakly coupled ones, these dualities are crucial to understand physics beyond perturbation theory. 

Recently,  much progress has been made through quantum information methods towards understanding the overarching principles behind dualities in the Hamiltonian framework. One key ingredient is the understanding that matter in gauge theories can be eliminated (syn. disentangled) from the Hamiltonian through exploitation of the Gauss law \cite{Haegeman_Gauging_Quantum_States_Global_Local_Symmetries,Zohar_2018, Zohar_2019, Tagliacozzo_2011}. Indeed, if one knows where Wilson lines end, one knows where matter resides. A second key ingredient is a procedure that lifts global symmetries to local (gauge) symmetries on the level of \emph{states}; at its core, this procedure amounts to adding a reference state in the pure gauge theory Hilbert space and projecting the combined state onto the gauge invariant subspace \cite{Haegeman_Gauging_Quantum_States_Global_Local_Symmetries}, and it turns out that this procedure is equivalent to the well known minimal coupling procedure in QFT. Combining these ideas allows one to take a pure matter theory and generate a dual theory in terms of gauge degrees of freedom through gauging (entangling), and then, upon switching the roles of matter and gauge fields, one may disentangle and obtain a "dual state".   On the lattice, a plethora of examples have been worked out (see, e.g. \cite{Ahskenazi_2022,Bravyi_2022,Tantivasadakarn_2024,Sukeno_2024}) and the general procedure is well understood \cite{Lootens_2024, PRXQuantum.4.020357}. Notably, this entire procedure can generally be represented by a finite depth quantum circuit \cite{PhysRevLett.134.130403} and it was proven that every 1+1 dimensional duality on the lattice can be understood through this procedure \cite{cuiper2025gaugingdualityonedimensionalquantum}.

In the continuum, and especially in the Hamiltonian framework, the relation between dualities and gauging has been less explored. In this letter, we wish to take a first step in that direction, by showing that two famous continuum dualities, T-duality and bosonization, can be recovered naturally in this framework. The upshot of this is a novel understanding of continuum dualities as well as the future potential to further learn from the many condensed matter results. 

For example, T-duality states that the 1+1 dimensional compact boson $\phi(x)$ at radius $R$ is dual to another compact boson $\tilde \phi(x)$ at radius $\tilde R = 1/(2\pi R)$ \cite{Alvarez:1996up,BUSCHER1988466}. In the Lagrangian framework these fields are related by $\partial_\mu \phi(x) = \epsilon_{\mu \nu} \partial^\nu \tilde \phi(x)$ and since $\partial_\mu \phi(x)$ is a conserved current, the temporal and spatial components of this equation respectively become Gauss and Ampere laws upon identification of the dual scalar $\tilde \phi(x)$ with the dimensionless electric field. Similarly, Abelian bosonization \cite{Fradkin_Field_Theories_of_Condensed_Matter_Physics,von_Delft_1998} relates fermions $\psi(x)$ to dual bosons $\phi(x)$ and is usually studied in the Hamiltonian framework. Among other expressions it is commonly stated that $\bar \psi(x) \gamma_0 \psi(x) = \partial_x \phi(x)$ which again becomes a Gauss law upon identification of the dual boson with the dimensionless electric field.

The outline of this paper is at follows. We start by introducing the general procedure directly in the continuum, closely following the known lattice constructions proposed in \cite{Haegeman_Gauging_Quantum_States_Global_Local_Symmetries,Zohar_2019,Ahskenazi_2022, Zohar_2018,Zohar_2019}. Then, we demonstrate that T-duality naturally fits in this framework. We discuss Abelian bosonization and find that the Chiral anomaly of the current algebra plays a crucial role in the construction of the disentangler. Finally, we conclude and discuss future directions.% In particular, we shortly mention a recent development where emergent spacetime appears through sequential gauging \cite{rubio2024emergent21dtopologicalorders}. 
\\
\\

\emph{Gauging and disentangling to get dualities. --}
Let $H$ be the Hamiltonian for a 1+1 dimensional quantum field theory with Fock space $\mathcal{H}_{matter}$. Let us assume the existence of a global $U(1)$ symmetry generated by charges $Q = \int dx Q(x)$ where $Q(x)$ is the local charge density. Since $H$ generates the time evolution, one can state that all the properties of the theory follow from its spectral properties. Therefore, a duality can be seen as an isometry between fock spaces $\mathcal{H}_{matter} \rightarrow \mathcal{H}_{dual}$ and corresponding operators $H \rightarrow H_{dual}$ constructed so that $H$ and $H_{dual}$ have identical spectral properties.

As mentioned in the introduction, the first step we use to create a duality is to lift the global $U(1)$ symmetry to a local (gauge) one, through the introduction of new, gauge degrees of freedom that will eventually represent the dual Hilbert space $\mathcal{H}_{gauge}=\mathcal{H}_{dual}$. More formally, define the electric field $E(x)$ and gauge field connection $A(x)$ satisfying $[A(x), E(y)] = -i\delta(x-y)$ \footnote{As is customary in the Hamiltonian formulation of gauge theories we have chosen a gauge where $A_0$ vanishes. }. With these definitions, we can define \emph{gauged states} 
\begin{align}
    \mathcal{G}\ket{\psi} = \frac{1}{2\pi }\int D \alpha \,  e^{i \int dx \, \alpha(x) G(x)}\ket{\psi} \otimes \ket{in}
\end{align}
where $\ket{in} = \ket{A=0}$ is the eigenvalue zero eigenstate of all $A(x)$ (i.e. it is a field coherent state), $\int D\alpha$ the functional integral and 
\begin{align}
G(x) = Q(x)- \frac{1}{g} \partial_x E(x)
\end{align}
with $g$ the gauge coupling generates gauge transformations. By construction, the gauged state $\mathcal{G}\ket{\psi}$ now satisfies the Gauss law $ G(x) \mathcal{G} \ket{\psi} = 0 $ and has therefore been supplemented with the correct Wilson lines. The prefactor $\frac{1}{2\pi}$ has been chosen so that the norm of $G \ket{\psi}$ is equal to that of $\ket{\psi}$. Similarly, for operators one defines
\begin{widetext}
\begin{align}
    \mathcal{G}[O] = \int D\alpha \,  e^{i \int dx \alpha(x) G(x) } \del{O \,  \otimes \ket{in}\bra{in}} e^{-i \int dx \alpha(x) G(x) }   
\end{align}
\end{widetext}
and it is easy to verify that
\begin{align}
	\mathcal{G} O \ket{\psi} = \mathcal{G}[O] \mathcal{G} \ket{\psi}
\end{align}
so that $O$ and $\mathcal{G}[O]$ have the same spectrum. In practice this means that $\mathcal{G}[O]$ is just the minimally coupled version of $O$ \cite{Haegeman_Gauging_Quantum_States_Global_Local_Symmetries,Zohar_2019,Ahskenazi_2022}. We note that this step will be identical for all examples below and we will not explicitly repeat it. 

The second step of the procedure is to disentangle the original matter degrees of freedom. To achieve this, we exploit the Gauss law to define a unitary operator $U_D$ that lowers the charge density $Q(x)$ by $g^{-1} \partial_x E(x)$ everywhere. With this, 
\begin{align}
	U_D \mathcal{G} \ket{\psi} = \ket{out} \otimes \ket{\tilde \psi} 
	\label{eq:magic of disentangler}
\end{align}
where $\ket{out} = \ket{Q = 0}$ is the eigenvalue zero eigenvector of all $Q(x)$ and $\ket{\tilde \psi}$ only lives in the gauge field Fock space $\mathcal{H}_{gauss}$. Since the disentangler $U_D$ is unitary, it is guaranteed that $U_D \mathcal{G}[O] U_D^\dagger$ and $\mathcal{G}[O]$ have the same spectrum and since $U_D \mathcal{G}[O] U_D^\dagger$ acts trivially on the matter degrees freedom the dual operator
\begin{align}
	\mathcal{UG}[O] = \braket{out | U_D \mathcal{G}[O] U_D^\dagger | out}
\end{align}	
has the same spectrum as $O$ up to some degeneracies on which the original operator $O$ acted trivially \cite{Zohar_2018,Zohar_2019}. For now, the existence of the disentangler will be assumed but in the upcoming text we will explicitly construct it for the cases of interest. Finally, we note that $\mathcal{UG}[O_A O_B] = \mathcal{UG}[O_A] \mathcal{UG}[O_B]$, this implies that operator algebras remain unchanged throughout duality transformations. 
\\
\\
\emph{T-duality. --}
For the Hamiltonian of the free compact scalar
\begin{align}
    H = \int_{-\infty}^\infty dx \del{ \frac{1}{2\beta^2} \pi(x)^2 + \frac{\beta^2}{2} (\partial_x \phi(x))^2 }
\end{align}
the relevant global $U(1)$ symmetry is has charge density $Q(x) = \pi(x)$ and one can show that 
\begin{align}
[\pi(y), e^{i \kappa \phi(x)}   ] = -\kappa e^{i \kappa \phi(x)} \delta(x-y)
\end{align}
 so that the vertex operators $e^{i \kappa \phi(x)}$ act as raising/lowering operators. Using this, one can easily check that the disentangler
\begin{align}
    U_D = e^{i \int dx \, \phi(x) g^{-1} \partial_x E(x) }
\end{align}
has all the desired properties (i.e. it is unitary and it lowers the charge density of gauge invariant states by $g^{-1}\partial_x E(x)$ everywhere). 

It is now easy to check that
\begin{align}
	\mathcal{UG}[e^{i \phi(x)}] = e^{i g \int_{-\infty}^x dx' A(x') }
\end{align}
i.e. the dual of a vertex operator is a half-infinite Wilson line. Next, let us find the dual Hamiltonian. Before gauging, we rewrite the Hamiltonian in terms of vertex operators so that it is clear where to attach Wilson lines. Additionally, we replace the derivatives with finite differences which gives rise to
\begin{widetext}
    \begin{align}
        %    H &= \int dx \del{ \frac{1}{2\beta^2} \pi(x)^2 + \frac{\beta^2}{2} \lim_{\epsilon \rightarrow 0} \frac{ e^{-i\phi(x)} - e^{-i\phi(x-\epsilon)}  }{\epsilon} \frac{e^{i\phi(x)} - e^{i\phi(x-\epsilon)}  }{\epsilon}  } \\
            H &= \int dx \Big( \frac{1}{2\beta^2} \pi(x)^2 + \lim_{\Delta \rightarrow 0} \frac{\beta^2}{2 \Delta^2 } \del{2 - e^{i \phi(x-\Delta)}e^{-i \phi(x)}  - e^{-i \phi(x-\Delta)}e^{i \phi(x)}    }   \Big) \nonumber \\
        \end{align}        
so that 
\begin{align}
    \mathcal{G}[H] = \int dx \del{ \frac{1}{2\beta^2} \pi(x)^2 - \lim_{\Delta \rightarrow 0} \frac{\beta^2}{2 \Delta^2 } \del{e^{i \phi(x-\Delta)} e^{ig \int_{x-\Delta}^x A(x') dx'} e^{-i \phi(x)}  + h.c.    }       }
\end{align}
\end{widetext}
where we dropped a constant. Finally, acting with the disentangler, projecting onto the charge zero sector and taking $\Delta \rightarrow 0$ gives:
% we finally obtain: 
%\begin{align}
%    U_D \mathcal{G}[H] U_D^\dagger = \int dx \del{ \frac{1}{2\beta^2} \del{\pi(x) + g^{-1}\partial_x E(x)  }^2 - \lim_{\epsilon \rightarrow 0} \frac{\beta^2}{2 \epsilon^2 } \del{ W_{x-\epsilon \rightarrow x} + W_{x-\epsilon \rightarrow x}^\dagger    }       }
%\end{align}
%where we have ignored an infinite constant which will shift the groundstate energy but not affect the dynamics. Finally, projecting onto the zero charge sector and expanding to leading order in $\epsilon$ leads to: 
\begin{align}
    \mathcal{UG}[H] = \int dx \del{ \frac{1}{2\beta^2} (g^{-1}\partial_x E(x))^2 + \frac{\beta^2}{2} ( g A(x))^2 } \ \ , 
\end{align}
where we have exploited the fact that the disentangler can also be interpreted as the transformation that performs a controlled shift of $\pi(x)$ by $g^{-1}\partial_x E(x)$. This result is now exactly the T-dual of the original Hamiltonian upon the identification of $\tilde \phi(x) = g^{-1} E(x)$ and $\tilde \pi(x) = g A(x)$. Crucially, this identification only works because the 1+1 dimensional scalar field has the same degrees of freedom as the 1+1 dimensional gauge field in the Weyl gauge. In this regard the self duality of the compact scalar is quite accidental for 1+1 dimensions.
\\
\\
\emph{Abelian bosonization. --}
The Hamiltonian of the free massless fermion is:
\begin{align}
    H = \int dx \, \bar \psi(x) \gamma_x (-i\partial_x) \psi(x)
\end{align}
where $\psi(x)$ is a two component spinor and $\bar \psi(x) = \psi^\dagger(x) \gamma_0$. As is well known, this Hamiltonian has a $(U(1) \otimes U(1))/Z_2 $ symmetry generated by the charges
\begin{align}
    Q_v &= \int dx \, :\bar \psi(x) \gamma_0 \psi(x) : \\
    Q_a &= \int dx \, :\bar \psi(x) \gamma_0 \gamma_5 \psi(x) :
\end{align}
where `` $:$ '' denotes the normal ordering with respect to the Dirac sea. Crucially, this normal ordering induces the anomalous current algebra
\begin{align}
	[Q_v(x), Q_a(x) ] = \frac{i}{\pi} \partial_x \delta(x-y) = -\frac{i}{\pi} \partial_y \delta(x-y)
	\label{eq:anomalous current commutator}
\end{align}
\cite{Vladimirov:1989ep, Mattis2004, Sykora:1999jd, PhysRev.177.2426,Bell:1969ts}. 

One can now proceed by gauging the vector or axial current but in what follows we will assume that the vector symmetry has been gauged. Using this gauged vector charge we define the disentangler 
\begin{align}
	U_D = e^{i \int dx \, Q_{\int a}(x) \pi g^{-1} \partial_x E(x) }
\end{align}
where we defined the \emph{partially integrated axial charge}
\begin{align}
	 Q_{\int a}(x) = \int_{-\infty}^x dx' \, Q_{a}(x')
\end{align}
that satisfies 
\begin{align}
	[Q_v(x), Q_{\int a}(y) ] = \frac{i}{\pi} \delta(x-y)
	\label{eq:integrated anomalous current commutator}
\end{align}
and therefore acts as a shift operator for the vector charge. Note that the anomaly is directly responsible for the fact that $:Q_{\int a}(x):$ acts as a shift operator. Therefore, as mentioned in the introduction, the anomalous current algebra played a crucial role in defining the disentangler and corresponding duality . Additionally, we note that $\ket{out} = \ket{ :Q_v(x): = 0}$ i.e. the state with zero vector charge everywhere. Again, due to the anomalous current algebra this state is unique, in particular we cannot additionally set $:Q_a(x):$ to zero everywhere.

We are now ready to find some dual operators. As a warmup, it is easy to see that:
\begin{align}
	\mathcal{UG}[:Q_v(x):] &= \braket{ out | U_D  :Q_v(x): U_D^\dagger | out } \\
	&= \braket{ out |   :Q_v(x): + g^{-1}\partial_x E(x)  | out } \\
	&=  g^{-1} \partial_x E(x)
\end{align}
where I used the facts that the disentangler acts as a shift operator for $:Q_v(x):$ and that $\ket{out}$ has zero vector charge density. Next, let us look at the dual of the axial current:
\begin{align}
	\mathcal{UG}[Q_a(x)] &= \braket{ out | U_D  \del{ Q_a(x) - \frac{g}{\pi} A(x) } U_D^\dagger | out }  \\
	&= \braket{ out  | U_D \del{- \frac{g}{\pi}  A(x)} U_D^\dagger | out } \\
	& = -\frac{g}{\pi}  A(x)
\end{align}
where we used the fact that the disentangler can also be interpreted as a shift operator for $A(x)$. Next, let us look at the chiral components of the fermion i.e. $\psi_{L,R}(x) = \frac{1 \pm \gamma_5}{2} \psi(x)$. We find 
\begin{widetext}
\begin{align}
 	\mathcal{UG}[\psi_{L,R}(x)] &=  \braket{ out | U_D  e^{i \int^x dx' g A(x') }  \psi_{L,R}(x) U_D^\dagger | out }  \\
	&= \braket{ out | e^{i \int^x dx' g \del{A(x') + \pi g^{-1} Q_a(x') } }  e^{\pm i \pi g^{-1} E(x) } \psi_{L,R}(x) | out }  \\
	&= \frac{1}{\sqrt{2 \pi \epsilon}} \, e^{i g \int^x dx' A(x') \pm i \pi g^{-1} E(x) } \label{eq : dual fermion}
 \end{align}
\end{widetext}

where we used \begin{align}
 	 \frac{1}{\sqrt{2 \pi \epsilon}}  = \braket{ out | e^{i \int^x dx' \pi Q_a(x') }  \psi_{L,R}(x) |out } \label{eq:prefactor}
 \end{align}
in which $\epsilon$ is the UV-regulator that appears in the properly regularized Fourrier expansion of the fermion field i.e. 
 \begin{align}
  \psi_{L,R}(x) = \int \frac{d k}{2 \pi} e^{i k x} \psi_{L,R}(k) e^{-\epsilon k/2} \label{eq:regularized fermion}\ \ .
 \end{align}

To prove \ref{eq:prefactor} it suffices to realise that $e^{i \int^x dx' \pi Q_a(x') }$ and $\psi_{L,R}(x) $ both lower the vector charge density by 1. Since $e^{i \int^x dx' \pi Q_a(x') }$ has operator norm 1, the right hand side of \ref{eq:prefactor} must therefore be the operator norm of $\psi_{L,R}(x)$. Using \ref{eq:regularized fermion} it is easy to see that $\{\psi^\dagger(x)_{L,R}, \psi(y)_{L,R} \} = \frac{1}{\pi} \frac{\epsilon}{\epsilon^2 + (x-y)^2}$ so that this operator norm is exactly the proposed $\frac{1}{\sqrt{2 \pi \epsilon}}$.

Finally, let us compute the dual of Hamiltonian, here it will be useful exploit $\mathcal{UG}[A B] = \mathcal{UG}[A] \mathcal{UG}[B]$. Furthermore, we will replace the derivative with a finite difference with this we obtain:
\begin{widetext}
\begin{align}
	\mathcal{UG}[H] &= \int dx \frac{1}{2\pi \epsilon}\frac{i}{2\Delta} \del{ e^{i g \int^x dx' A(x') + i \frac{\pi}{g}E(x) }  e^{-i g \int^{x+\Delta} dx' A(x') - i \frac{\pi}{g}E(x+\Delta) } + h.c. + L \leftrightarrow R} \\ 
	&= \int dx \frac{1}{2\pi \epsilon}\frac{i}{2\Delta} \del{ i e^{ -i \sqrt{\pi}\del{ \frac{g}{\sqrt{\pi}} \int_x^{x+\Delta} dx' A(x')  + \frac{\sqrt \pi}{g} \del{E(x+\Delta) -E(x)}  } } + h.c. + L \leftrightarrow R} \label{eq:half_dual}
\end{align}
\end{widetext}

where the second factor $i$ on the second line has arised due to careful application of the Baker–Campbell–Hausdorff formula. To proceed we wish to take the limit $\Delta \rightarrow 0$ but care is required because (\ref{eq:regularized fermion}) implies that fluctuations of the fermion-field (as well as its dual) are not defined on length scales smaller than $\epsilon$. Therefore we must always keep $\Delta \gg \epsilon$. Additionally, we cannot just expand the exponential of \ref{eq:half_dual} because $A(x)$ and $E(x)$ are both unbounded operators. To deal with this, we must normal order the exponentials with respect to some reference state (the empty state) and then expand the obtained expression. Physically this means that we are expanding the exponential under the assumption that it acts on states that are sufficiently close to the dual of the Dirac sea. After the lengthy derivation of appendix A this leads to: 
\begin{widetext}
\begin{align}
	\mathcal{UG}[H] &= \int dx \frac{-1}{2\pi \epsilon}\frac{1}{2\Delta} \del{ : e^{ -i \sqrt{\pi}\del{ \frac{g}{\sqrt{\pi}} \int_x^{x+\Delta} dx' A(x')  + \frac{\sqrt \pi}{g} \del{E(x+\Delta) -E(x)}  } } : \sqrt{\frac{\epsilon^2}{\epsilon^2 + \Delta^2}}+ h.c. + L \leftrightarrow R} 
\end{align}
\end{widetext}
which can now safely be expanded in $\Delta$. Finally, taking the limit $\Delta \rightarrow 0$ while keeping $\Delta \gg \epsilon$ leads to
\begin{align}
	\mathcal{UG}[H] =  \frac{1}{2} \int dx: \frac{\pi}{g^2}\del{\partial_x E(x) }^2 + \frac{g^2}{\pi} A^2(x) : - \int dx \frac{1}{\pi \Delta}
\end{align}
where the second divergent term can be interpreted as the divergent energy of the Dirac sea. Ignoring this, which amounts to normal ordering the original fermionic Hamiltonian leads to the compact boson Hamiltonian upon identification of $\phi(x) = \frac{\sqrt \pi}{g} E(x)$ and $\pi(x) = \frac{g}{\sqrt \pi} A(x)$. 
\\
At this point, it is interesting to comment on the subtlety that the boson is only dual to the fermion when the latter is coupled to a $Z_2$ gauge symmetry $\psi(x) \rightarrow -\psi(x)$ that eliminates the single fermion from the spectrum. Here this result naturally arises from the underlying physics. Indeed, since the original fermions are interpreted as endpoints of Wilson lines we cannot use local operators in the bosonic theory to make a single fermion. In fact, one can see that the original fermion is mapped to an operator that changes the boundary conditions of the bosonic theory. Imposing fixed boundary conditions in the bosonic theory naturally leads to the gauge constraint on the fermionic Fock space.   \\
\\
\emph{Conclusion and outlook. --}
We have demonstrated that the gauging and disentangling procedure that formalizes the concept of dualities on the lattice \cite{Lootens_2024,PRXQuantum.4.020357,PhysRevLett.134.130403} can be expanded to the continuum. As a benchmark we demonstrated that this allows one to rederive T-duality and bosonization in a new way that is transparent as well as mathematically rigorous. Notably, the derived dualities are isometries between Fock spaces that do not rely on any properties of the involved Hamiltonians other than the conservation of particle number. Therefore, our derivation remains valid for most interacting theories as well as theories defined on curved backgrounds that conserve particle number. Furthermore, this method can also be applied to T-duality on a finite system with Neumann or Dirichlet boundary conditions. Here the duality can still be constructed but turns out to switch the boundary conditions, this is consistent with what is known in string theory \cite{Alvarez:1996up}.

Similarly, it is straightforward to reproduce non-Abelian bosonization using this framework. Here the only caveat is that one should gauge the $U(1)^N$ subgroup of vector symmetry rather than the full  $SU(N)$ vector symmetry. Indeed, only the former is sufficient to fully disentangle the original matter degrees of freedom in a way similar to the one proposed above. Doing so leads to a dual theory in terms of $N$ $U(1)$ gauge fields and one can recombine these into group elements of $SU(N)$ using the prescription given in \cite{doAmaral:1991em}.

Notably, this work lays the foundation for further crossover of many recently developed lattice approaches to dualities directly in the continuum. For example in \cite{rubio2024emergent21dtopologicalorders} it was noticed that gauged operators and states have a new global symmetry that acts only on the gauge fields. One can then also gauge this global symmetry and thereby introduce another set of gauge fields that have another global symmetry. Sequential application of this procedure leads to an emergent higher dimensional space and it is understood that the emergent bulk state is topologically ordered. Can such a construction be done in the continuum? If so, the resulting bulk state will most likely be of the form proposed in \cite{Roose:2025ibv}. In particular we are currently trying to connect WZW and CS in this framework. One might be optimistic and recall Ads/CFT\cite{maldacena2004tasi2003lecturesadscft}, especially since 2+1 dimensional gravity is understood to be topologically ordered \cite{WITTEN1988601}. 

Additionally, taking the works \cite{Lootens_2024,PRXQuantum.4.020357,PhysRevLett.134.130403} into the continuum allows us to study the framework in the presence of a nontrivial static background metric. 

\emph{acknowledgements. --}
The authors acknowledge valuable discussions with Laurens Lootens, Frank Verstraete, Ignacio Cirac, Jose Garre Rubio, Amit Sever,  Alfred Benedito and Itay Gomelski. This research is funded by the European Union (ERC, OverSign, 101122583). Views and opinions expressed are however those of the author(s) only and do not necessarily reflect those of the European Union or the European Research Council. Neither the European Union nor the granting authority can be held responsible for them. We acknowledge the support of the Israel Science Foundation (Grant No. 374/24). 

\bibliography{biblio}

\appendix
\section{APPENDIX
A} \label{sec : appendix A}
In this appendix we prove that
\begin{align}
	A &= e^{ -i \sqrt{\pi}\del{ \frac{g}{\sqrt{\pi}} \int_x^{x+\Delta} dx' A(x')  + \frac{\sqrt \pi}{g} \del{E(x+\Delta) -E(x)}  } }  \\ 
	&= :A: \sqrt{\frac{\epsilon^2}{\epsilon^2 + \Delta^2}}
\end{align} 

To start, we expand the relevant fields in terms of creation and annihilation operators i.e.
\begin{align}
	\frac{\sqrt \pi}{g} E(x) &= \int \frac{dk}{2 \pi} \frac{1}{\sqrt{2 |k|}} e^{- \frac{\epsilon |k|}{2}} \del{e^{i k x } a(k) + h.c.  } \\
	\frac{g}{\sqrt \pi} A(x) &= \int \frac{dk}{2 \pi} \sqrt{\frac{|k|}{2}} e^{- \frac{\epsilon |k|}{2}}   \del{ie^{i k x } a(k) + h.c.  } 
\end{align}
where $[a(k), a^\dagger(l) ] = 2 \pi \delta(k-l)$. With this, 
\begin{align}
	A = e^{-2 i \sqrt{\pi} \int_0^\infty \frac{dk}{2\pi} \frac{1}{\sqrt{2 k}} \del{ \del{e^{ik(x+\Delta) - e^{ikx}}}a(k) +h.c. } }
\end{align}
and using the Zassenhaus formula this can be rewritten as
\begin{align}
	A = :A: e^{-\pi \mathcal{N}} \label{eq:intermediate}
\end{align}
where the normal ordered operator is
\begin{widetext}
\begin{align}
	:A: = e^{-2 i \sqrt{\pi} \int_0^\infty \frac{dk}{2\pi} \frac{1}{\sqrt{2 k}} \del{ e^{-ik(x+\Delta) - e^{-ikx}}}a^\dagger(k)  }e^{-2 i \sqrt{\pi} \int_0^\infty \frac{dk}{2\pi} \frac{1}{\sqrt{2 k}} \del{ e^{ik(x+\Delta) - e^{ikx}}}a(k)  }
\end{align}
and
\begin{align}
	\mathcal{N} &= \left[   \int_0^\infty \frac{dk}{2 \pi} \frac{1}{\sqrt{k}} e^{-\frac{\epsilon k }{2}}\del{e^{-ik(x+\Delta)} - e^{-ikx}} a^\dagger(k)  ,       \int_0^\infty \frac{dl}{2 \pi} \frac{1}{\sqrt{l}} e^{-\frac{\epsilon l }{2}}\del{e^{il(x+\Delta)} - e^{ilx}}   a(l)    \right] \ \ .
\end{align}
\end{widetext}
After some easy manipulations we find
\begin{align}
	\mathcal{N} &= \int_0^\infty \frac{dk}{2\pi}\frac{1}{k}e^{-\epsilon k }2(1-cos(k \Delta)) \\
	&= I(\Delta) + I(-\Delta) 
\end{align}
where 
\begin{align}
	I(\Delta) = \int_0^\infty \frac{dk}{2\pi}\frac{1}{k}e^{-\epsilon k } (1-e^{i k \Delta})
\end{align}

To compute $I(\Delta)$ we note
\begin{align}
	\frac{\partial I(\Delta)}{\partial \Delta} &= -i \int_0^\infty \frac{dk}{2\pi} e^{-\epsilon k} e^{i k \Delta} \\
	&= \frac{i}{2\pi}\frac{1}{i \Delta - \epsilon}
\end{align}
so that 
\begin{align}
	I(\Delta) - I(0) &= \frac{i}{2\pi}\int_0^\Delta d\Delta' \frac{1}{i \Delta' - \epsilon} \\
	&=\frac{1}{2\pi}	\log\del{\frac{\epsilon - i \Delta}{\epsilon}}  \ \ .
\end{align}
Since $I(0) = 0$ this leads to 
\begin{align}
	\mathcal{N} = \frac{1}{2\pi}\log\del{\frac{\epsilon^2 + \Delta^2}{\epsilon^2}}
\end{align}
which can be plugged into \ref{eq:intermediate} to get the desired result.

\end{document}